\documentclass[
superscriptaddress,
nofootinbib,
 amsmath,amssymb,
 aps,
floatfix,
prd,
showkeys,
10pt,
twocolumn,
]{revtex4-2}

\usepackage{xcolor}

\usepackage{bbm}
\usepackage{ulem}
\usepackage{tabularx}
\usepackage{graphicx}
\usepackage{amsmath}
\usepackage{hyperref}
\hypersetup{
    colorlinks,
    linkcolor={red!50!black},
    citecolor={blue!50!black},
    urlcolor={blue!80!black}
}

\graphicspath{{./}{figures/}}

\usepackage{amsfonts,amssymb,stmaryrd,latexsym,amsmath,braket}
\usepackage{graphicx,subfigure}
\usepackage{times}
\usepackage{slashed}
\usepackage{braket}
\usepackage{verbatim}
\usepackage[utf8]{inputenc}

\usepackage{color}
\usepackage[dvipsnames]{xcolor}

\usepackage{hyperref}
\usepackage{bm}
\usepackage{soul}
\usepackage{ulem}

\begin{document}

\title{Information content of the Gamow Teller Resonance:  Bayesian inference with self-consistent Skyrme models}

\author{Zidu Lin}
\affiliation{Sino-French Institute of Nuclear Engineering and Technology, Sun Yat-Sen University, Zhuhai 519082, China}
\affiliation{Department of Physics and Astronomy, University of Tennessee Knoxville, Knoxville, TN 37996-1200, USA.}
\affiliation{Department of Physics and Astronomy, University of New Hampshire, Durham, NH 03824, USA.}

\author{Gianluca Col\`o}
\affiliation{Dipartimento di Fisica, Universit\`a
degli Studi di Milano, via Celoria 16, 20133 Milano, Italy,}
\affiliation{INFN, Sezione di Milano, via Celoria 16, 20133 Milano, Italy}

\author{Andrew W. Steiner}
\affiliation{Eureka Scientific, Inc., 2452 Delmer Street, Oakland, CA 94602, USA}
\affiliation{Department of Physics and Astronomy, University of Tennessee Knoxville, Knoxville, TN 37996-1200, USA.}
\affiliation{Physics Division, Oak Ridge National Laboratory,TN 37831, USA.}

\author{Amber Stinson}
\affiliation{Department of Physics and Astronomy, University of Tennessee Knoxville, Knoxville, TN 37996-1200, USA.}

\begin{abstract}

Many important charge-exchange weak reactions (e.g. beta decays, double beta decays, neutrino-nucleus reactions and neutrino-nucleon reactions) are closely related to the spin- and isospin-dependent nuclear force. However, few of these reactions can be precisely measured (or even observed) by current terrestrial experiments. The Gamow-Teller resonance (GTR) in finite nuclei is one of the few experimental evidences that reveals the features of the spin- and isospin-dependent force. In Lin {\it et al.}  \cite{Lin:2025glw}, we obtain the probability distribution of the Landau-Migdal parameter $G'_0$ (that is a parameter characterizing the main part of the spin-isospin force) from measurements of the Gamow-Teller resonance by using Bayesian inference. In this follow-up work, the correlations emerging from our Bayesian analysis are thoroughly studied in a quantitative manner.
Analysis of the Bayesian posterior gives two standard Skyrme parameterizations (the nuSky* model) that successfully reproduce, in a unified manner, ground‑state observables of finite nuclei, the salient features of the GTR, and the bulk properties of neutron‑star matter.
Since the collective excitations in finite nuclei and those in neutron‑star matter are related, we constrain the axial‑current, charged‑current neutrino–nucleon reaction rates in hot and dense compact‑star matter using the Skyrme interactions sampled from the Bayesian posterior, as the first astrophysical application of this Bayesian study.
The GTR-informed calculations efficiently diminish the spread in the values of the neutrino opacity; nevertheless, significant 
uncertainties on this quantity remain,
due to the not-so-well-constrained symmetry energy at sub-saturation density. 
Fixing this quantity on top of a well-constrained spin-isospin interaction would
pave the way towards a systematic determination of neutrino-dense matter interactions in the future.     
\end{abstract}

\maketitle

\section{Introduction.} 

Studying the ground-state and excited-state properties of finite nuclei reveals the fundamental nature of nuclear structures and nucleon-nucleon interactions. By applying these laboratory-tested interactions to astrophysics, we can better understand the dense matter phases and neutrino reaction rates that drive compact star evolution.

Over the past several decades, substantial effort has been devoted to developing self‑consistent and accurate descriptions of global nuclear properties such as masses and charge radii as well as other observables across the nuclear chart. 
Several 
Skyrme, Gogny and relativistic energy functionals~\cite{Vautherin1972, chabanat1997, Decharge1980, Berger1991, Lalazissis, Meng2006, 
Kortelainen:2010hv,
Zhao2010, 
Kortelainen:2013faa,
Chen:2014mza, 
Zhao2022}
have been constructed to achieve this goal and are now widely used to investigate the structure of exotic nuclei and nucleosynthesis pathways. With the rapid improvements of nuclear experiments \cite{Wakasa:2012zz,PhysRevC.52.604,PhysRevLett.121.132501,PhysRevC.55.2909,PhysRevLett.103.012503,PREX:2021umo,CREX:2022kgg}, more interesting nuclear properties (e.g. the Gamow-Teller resonance (GTR) and the neutron skin thickness) have been measured. A key question is whether current nuclear structure theories can simultaneously account for these new observables alongside fundamental ground-state properties, such as charge radii and masses. Consequently, these measurements present both a challenge and an opportunity to refine our understanding of nuclear forces, structure, and the nature of compact stars.

In this work, we first systematically investigate the capability of Skyrme models to self-consistently describe both the basic ground-state properties and the features of the GTR spectrum of $^{90}$Zr, $^{132}$Sn and $^{208}$Pb, by analyzing their parameters and associated correlations using state-of-the-art Bayesian inference. As a result, we obtain a large ensemble of Skyrme interactions (denoted GTAll\cite{zlin2025zenodo}), sampled from the Bayesian posterior, that simultaneously reproduce the ground‑state observables and GTR properties of these nuclei within reasonable uncertainties. 

Collective excitations arise not only in finite nuclei but also in uniform matter encountered in compact stars. When neutrinos traverse the hot and dense nuclear medium present in core‑collapse supernovae and binary‑neutron‑star mergers, collective modes in the Gamow–Teller channel can be excited; this mode is directly analogous to the Gamow–Teller resonance observed in finite nuclei \cite{Burrows:1998cg}. Consequently, as an astrophysical application of the GTAll ensemble which is constrained jointly by ground‑state observables and GTR data, we employ these Skyrme energy density functionals to predict the charged‑current axial‑current neutrino–nucleon opacities in hot and dense matter relevant to the evolution of neutron‑star mergers and core‑collapse supernovae.

The paper is organized as follows. Section II introduces the random phase approximation (RPA) for describing GTR spectra in finite nuclei, the Bayesian inference framework used to constrain the Skyrme interactions, and the RPA‑based calculation of neutrino–matter reaction rates in compact‑star environments. Section III presents the correlations emerging from the Bayesian posterior and identifies two Skyrme parameterizations that successfully reproduce selected ground‑state observables, GTR features, and basic neutron‑star properties. Finally, we present the probability density distribution of charged‑current axial‑current neutrino opacities obtained using the GTAll ensemble of Skyrme parameterizations.

\section{Methods} 

In this section, the RPA method for describing the GTR in finite nuclei, the Bayesian inference of Skyrme functionals, and the RPA method for describing the response properties of infinite nucleonic matter, induced by the perturbation of strong neutrino fluxes in compact stars, are introduced.    
\subsection{RPA for finite nuclei }

The RPA method in this work is the same as in Ref. \cite{Lin:2025glw}. For completeness, we show the key equations of this method below.
The RPA matrix equation is
\begin{equation}
    \begin{pmatrix}
        A & B \\
        -B^* & -A^*  
    \end{pmatrix}\begin{pmatrix}
        X^\nu \\
        Y^\nu
    \end{pmatrix}=E_\nu\begin{pmatrix}
        X^\nu \\
        Y^\mu
    \end{pmatrix}~,
\end{equation}
on a particle-hole basis.
The matrix elements $A$ and $B$ are defined as
\begin{equation}\label{eq:rpamatrixelement}
    A_{mi,nj}=(\epsilon_m-\epsilon_i)\delta_{mn}\delta_{ij}+\langle mj|V_\mathrm{res}|in\rangle,
\end{equation}
\begin{equation}\label{eq:rpamatrixelement2}
    B_{mi,nj}=\langle mn|V_\mathrm{res}|ij\rangle,
\end{equation}
where occupied (unoccupied) states are denoted as $i, j$ ($m,n$). The $X^\nu$ and $Y^\nu$ are the amplitudes of the eigenvectors of the RPA matrix and $E_n$ are the corresponding eigenvalues. 
$V_\mathrm{res}$ is the residual interaction. 

The Gamow-Teller strength function 
is obtained from the eigenstates, labeled $|\nu\rangle$, as
\begin{equation}
    S(E)=\sum_\nu |\langle \nu||\hat{O}^{\pm}_\mathrm{GT}||0\rangle|^2\delta(E-E_\nu),
\end{equation}
where $\hat{O}^\pm_\mathrm{GT}=\sum_{i=1}^A \sigma^i_\pm \tau_\pm^i $. The eigenstate $|\nu\rangle$ is constructed by letting the excitation operator $O_\nu^+$ act on the RPA ground state $|\tilde0\rangle$:
\begin{equation}
    |\nu\rangle=O_\nu^+|\tilde0\rangle,
\end{equation}
and has the following structure:
\begin{equation}    O^+_\nu=\sum_{mi}X^\nu_{mi}Q^+_{mi}(JM)-Y^\nu_{mi}Q_{mi}(\widetilde {JM}),
\end{equation}
where the $\widetilde{JM}$ is the time-reversed state of $JM$.
The density-dependent Landau-Migdal parameter $G'_0(n)$ is included in the RPA matrix elements $A$ and $B$. %

\subsection{Bayesian Inference}

The intrinsic scattering (IS) parameters representing the systematic uncertainty of theoretical models are applied in the Bayesian inference of this work as additional degrees of freedom. The motivation, the physical meaning of these IS parameters, and the prior distribution applied in this work are introduced in Ref. \cite{Lin:2025glw}. %

The posterior distribution of the Skyrme parameterizations is proportional to 
\begin{equation}\label{eq:postorig}
    P_{\mathrm{post}}(\{p\}) \propto {\cal{L}}(\{p\}) P_{\mathrm{prior}}(\{p\})~,
\end{equation}
where $\{p\}$ is the set of Skyrme parameters and ${\cal{L}}(\{p\})$ is the likelihood constraining the parameter distribution of our model. Given $P_{\mathrm{post}}(\{p\})$, we have the corresponding posterior distribution of $G'_0$, which is a function of $\{p\}$. The details on the definition of $G'_0$ and on the various conventions introduced in the literature for it, are given in our Ref.~\cite{Lin:2025glw}. 
Note that if we ignore the theoretical systematic uncertainty, the likelihood ${\cal{L}}(\{p\})$ can be written as
\begin{equation}
    {\cal{L}}(\{p\}) = \prod_k \exp \left\{ - \frac{\left[O^\mathrm{Th}_k(\{p\})-O_k\right]^2}{2 \sigma^2_k} \right\},
\end{equation}
where $O^\mathrm{Th}_k$ is the theoretical prediction of the observable $O_k$, and $\sigma_k$ is the experimental uncertainty of $O_k$.

However, in many cases, the systematic uncertainty coming from theoretical calculations can be  larger than the experimental uncertainties (e.g. in the case of calculating the nuclei binding energy and charge radii). This is of course related to the capability of the model to reproduce experimental data. In this work we consider the unknown systematic theoretical uncertainties in the likelihood by using ~\cite{Al-Mamun:2020vzu}
\begin{equation}\label{eq:islikelihood}
\begin{split}
    {\cal{L}}(\{p\}) &= \prod_k \frac{1}{N_k}\int_a^b \exp \left\{ - \frac{\left[O^\mathrm{Th}_k(\{p\})-\hat{O}_k\right]^2}{2 \sigma^2_\mathrm{k,sys}} \right\}\\&\cdot\exp \left\{ - \frac{\left[O_k-\hat{O}_k\right]^2}{2 \sigma^2_k} \right\}d \hat{O}_k,
\end{split}
\end{equation}
where $\hat{O}_k$ is the ``probabilistic" output from the theory with the mean of $O_k^\mathrm{Th}$ and with the uncertainty of $\sigma_\mathrm{k,sys}$. Consequently, $\hat{O}_k$ can deviate from $O^\mathrm{Th}_k$ due to the theoretical uncertainty. Note that in Eq. (\ref{eq:islikelihood}) the normalization factors $N_k$ cannot be ignored since they are an implicit function of $\sigma_\mathrm{k,sys}$. Also note that, 

\begin{equation}\label{eq:islikelihood2}
\begin{split}
   \int_{-\infty}^\infty &\exp \left\{ - \frac{\left[O^\mathrm{Th}-\hat{O}\right]^2}{2 \sigma^2_{sys}} \right\}\cdot\exp \left\{ - \frac{\left[O_k-\hat{O}_k\right]^2}{2 \sigma^2} \right\}d \hat{O}\\&\propto \mathrm{exp} \left\{-\frac{(O^\mathrm{Th}-\hat{O})^2}{2(\sigma^2_{sys}+\sigma^2)}\right\},
\end{split}
\end{equation}
where the right-hand-side formula is commonly used in calculating the likelihood with model uncertainties. However, the predicted physical quantities have finite boundaries. For example, $m_0\%$ should be within the interval of [$0\%$,$100\%$] 
and $R_\mathrm{ch}$ should always be positive. The upper and lower bound of the integral in Eq.~(\ref{eq:islikelihood}) is chosen to make sure that $\hat{O}_k$ is always physical. The lower bound of the model predictions of average binding energy per nucleon BE/A, $E_\mathrm{GT}$, and spin-orbit splitting values is 0 MeV, and the upper bound of these quantities is set to be a large enough value, which is 40 MeV in this work. The lower bound of $R_\mathrm{ch}$ is 0 fm and the upper bound of $R_\mathrm{ch}$ is set to be a large enough value, which is 10 fm in this work. The upper and the lower bound of $m_0\%$ is $0\%$ and $100\%$ respectively.

Similar as in Ref. \cite{Lin:2025glw}, the Skyrme models are constrained by charge radii and binding energies ($^{208}$Pb, $^{132}$Sn, $^{90}$Zr and $^{48}$Ca), by the neutron skin thickness (of $^{208}$Pb and $^{48}$Ca), by 9 spin-orbit splitting values ($^{208}$Pb, $^{132}$Sn, $^{48}$Ca), by the GTR peak energy $E_\mathrm{GT}$ and the percentage of strength exhausted in the GTR region,  $m_0\%$ ($^{208}$Pb, $^{132}$Sn, $^{90}$Zr). The first Bayesian inference include all 25 constraints (denoted as GTAll). In addtion, we perform another Bayesian inference having all the constraints except for the GTR constraints, and is denoted as noGT.    

\subsection{RPA for hot and dense matter}

The double differential cross section per volume of neutrino-nucleon absorption reactions is written as \cite{Reddy:1997yr,Burrows:1998cg}: 

\begin{equation}\label{eq:neutrinombfull}
\begin{split}
    \frac{1}{V}\frac{d^2\sigma^\pm_{}}{dcos\theta dE_{e^-/e^+}}=&\frac{G^2_F\mathrm{cos}\theta^2_C p_{e^-/e^+}E_{e^-/e^+}}{4\pi^2}\times\\&[g^2_\mathrm{V}(1+\mathrm{cos}\theta)S_V(q_0,q)+\\&g^2_\mathrm{A}(3-\mathrm{cos}\theta)S_A(q_0,q)],
\end{split}
\end{equation}
where $G_\mathrm{F}$ is the Fermi weak coupling constant ($\approx 1.1663787 \times 10^{-5}\mathrm{GeV^{-2}}$), $\theta_\mathrm{C}$ is the Cabibbo angle ($\approx 14.9$ degrees), $g_\mathrm{V}=1$, $g_\mathrm{A}=1.27$, $S_{\mathrm{V/A}}$ stands for the structure factors of vector (axial-vector) channels of neutrino absorption reactions.  Note that, due to the $1+\mathrm{cos}\theta$ ($3-\mathrm{cos}\theta$) term in the vector (axial-vector) channel in Eq.~(\ref{eq:neutrinombfull}), a larger contribution from axial-current reactions to neutrino absorptions is expected. 

Here, we introduce the basic formulas for calculating the many-body correction on charged current neutrino-nucleon reactions based on RPA. Based on linear response theory \cite{Fetter1971QuantumTO}, the structure factor of neutrino-nucleon reactions is estimated as
\begin{equation}\label{eq:dynamicS}
    S(q_0,q)=\frac{2\mathrm{Im}\Pi}{1-\exp(-z)},
\end{equation}
where $z=(q_0+\mu_n-\mu_p)/T$ for the reaction $\nu_e+n\rightarrow p+e^-$. The $q_0$ and $q$ are the transferred energy and momentum of neutrino-nucleon reactions. %

In frequency space, the polarization function $\Pi$ is
\begin{equation}
    \Pi(q_0,q)=2\int \frac{d^3k}{(2\pi)^3} G(q_0,q,\Vec{k})\equiv2\langle G \rangle,
\end{equation}
where $G(q_0,q,\Vec{k})$ is a particle-hole propagator and $\vec{k}$ is the corresponding momentum of the hole. In this work, $G(q_0,q,\Vec{k})$ is approximated in terms of $G_{RPA}$, that obeys the Bethe-Salpeter equation, namely
\begin{equation}\label{eq:BS}
\begin{split}
    G^\alpha_{RPA}&(q_0,q,\Vec{k}_1)=G_{HF}(q_0,q,\Vec{k}_1)+G_{HF}(q_0,q,\Vec{k}_1)\\&\times\sum_{\alpha'}\int\frac{d^3k_2}{(2\pi)^3}V^{\alpha,\alpha'}_{ph}(q,\vec{k}_1,\vec{k}_2)\times G^{\alpha'}_{RPA}(q_0,q,\vec{k}_2),
\end{split}
\end{equation}
with $\alpha=0\ (1)$ representing non-spin-flip (spin-flip) channels and $V^{\alpha,\alpha'}_{ph}=\delta(\alpha-\alpha')[W_1^\alpha(q)+W_2^\alpha(q)(\Vec{k}_1-\Vec{k}_2)^2]$. Because of the Kronecker delta function $\delta(\alpha-\alpha')$, the spin-flip and non-spin-flip channels are not coupled to each other. The $W^\alpha_{1,2}$ are the residual interaction terms for the charge-exchange RPA in uniform matter and are functions of the Skyrme parameters 
as is written in detail in~\cite{Hernandez:1999zz,PhysRevC.107.015804}. Note that
\begin{equation} \label{eq:residual}
    G'_0=\frac{N_0}{2}(\frac{1}{2}W^1_1+W_2^1 k^2_\mathrm{F}),
\end{equation}
where $k_\mathrm{F}$ is the Fermi momentum  and $N_0=\frac{2m^*k_\mathrm{F}}{\pi^2}$.  It is then clear that $G'_0$ is related not only to the GTR spectrum of finite nuclei (where the proton fraction is close to or moderately smaller than 0.5), but also influences the  $S_\mathrm{A}$ modifying the axial-current contributions to the charged-current neutrino opacity in neutron-rich compact star matter.

$G_\mathrm{HF}$ is defined as
\begin{equation}
\label{eq:HFpropagator}
    G_\mathrm{HF}=\frac{n_n(\vec{k})-n_p(\vec{k}+\vec{q})}{q_0-(E_p(\vec{k}+\vec{q})-E_n(\vec{k}))+i\eta}.
\end{equation}
To solve the Bethe-Salpeter equation, we follow the strategy proposed in ~\cite{Hernandez:1999zz,Margueron:2006wh}. By integrating both the right- and left-hand sides of the equation  (\ref{eq:BS}) over $\Vec{k}_1$ and expanding in multipoles the residual interaction $V_{ph}$, the Bethe-Salpeter equation can be re-written as a system of algebraic equations, from which the $\langle G\rangle_{RPA}=\Pi/2$ can be solved for:
\begin{equation}
\begin{split}
   <&G>_\mathrm{RPA}=(\beta_0+2W_2^\alpha(\beta_0\beta_3-\beta_1^2)q^2)\\&/\{1-W_1^\alpha\beta_0+2q^2W_2^\alpha[-\beta_2+\beta_3+W_1^\alpha(\beta_1^2-\beta_0\beta_3)]\\&+q^4(W_2^\alpha)^2[-\beta_0\beta_5+4(\beta_1\beta_4-\beta_2\beta_3)+\beta_2^2]\\&+2q^6(W_2^\alpha)^3[-\beta_0\beta_3\beta_5+\beta_0\beta_4^2+\beta_1^2\beta_5-2\beta_1\beta_2\beta_4+\beta_2^2\beta_3]\},
\end{split}
\end{equation}
where $W^\alpha_i$ are given by Eqs.~(7-10) of Ref.~\cite{Hernandez:1999zz}, $\beta_l=\int\frac{d^3k}{(2\pi)^3}G_{HF}F_l$, $\langle G\rangle_\mathrm{RPA}=\int\frac{d^3k}{(2\pi)^3}G_{RPA}$ and 
\begin{equation}
    F_{0-5}=\{1,\frac{\Vec{k}\cdot\Vec{q}}{q^2},\frac{k^2}{q^2},\frac{(\Vec{k}\cdot\Vec{q})^2}{q^4},\frac{\Vec{k}\cdot\Vec{q} ~k^2}{q^4},\frac{k^4}{q^4}\}. 
\end{equation}

Given $\langle G\rangle_\mathrm{RPA}$, we can then obtain the dynamic structure factor $S(q_0,q)$ by using Eq. (\ref{eq:dynamicS}). Interested readers are referred to ~\cite{Hernandez:1999zz,PhysRevC.59.2888,PhysRevC.107.015804} for previous detailed calculations of the many-body correction on neutrino opacities in core-collapse supernvoae (CCSNe) and binary neutron star (BNS) mergers.

In Eq.~(\ref{eq:HFpropagator}), the dispersion relation of nucleons follows $E_\mathrm{N}(\vec{k})=\frac{k^2}{2m^*}+U_\mathrm{N}$. Since $n_N=2\int \frac{dk^3}{(2\pi)^2}\frac{1}{\mathrm{exp}[(E_N(k)-\mu_N)/T]+1}$, we can solve for $U_N$ given $n_N$, $T$, $m^*$ and $\mu_N$. 
Consequently, the in-medium potential difference $\delta U=U_\mathrm{n}-U_\mathrm{p}$ heavily influences the Hartree-Fock propagator $G_\mathrm{HF}$, as well as $G_\mathrm{RPA}$ and $S(q_0,q)$. 

The Skyrme parameterizations sampled from the GTAll posterior are then used to calculate many-body corrections on charged-current neutrino opacity, which play important roles in understanding the explosion mechanism of core-collapse  supernovae (CCSNe) and the formation of heavy elements in both CCSNe and binary neutron star (BNS) mergers ~\cite{Reddy:1997yr,PhysRevC.59.2888,Burrows:1998cg,PhysRevC.95.025801,PhysRevLett.132.232502,PhysRevC.107.015804}. To quantify the effect of many-body corrections on charged current neutrino-nucleon reactions, we define:
\begin{equation}\label{eq:neutrinomb}
\begin{split}
    \mathrm{MB}^{(\nu_{e}/\bar{\nu}_e,e^-/e^+)}_{ax} &=\frac{\int \frac{1}{V}\frac{d^2\sigma^\pm_{ax}}{dcos\theta dE_{e^-/e^+}}dcos\theta dE_{e/e^+}}{\int \frac{1}{V}\frac{d\sigma^\pm_{0,ax}}{dcos\theta}dcos\theta },
\end{split}
\end{equation}
where the integrand of the numerator is the axial-current differential  neutrino-nucleon cross section per volume including many-body corrections:
\begin{equation}\label{eq:neutrinomb2}
\begin{split}
    \frac{1}{V}\frac{d^2\sigma^\pm_{ax}}{dcos\theta dE_{e^-/e^+}} &=\frac{G^2_Fg^2_Ap_{e^-/e^+}E_{e^-/e^+}}{4\pi^2}\\&\times(3-cos\theta)S_A(q_0,q).
\end{split}
\end{equation}
The value of $g_A$ is $\approx1.27$. The many-body correction is encoded in $S_A(q_0,q)$ derived in Eq.~(\ref{eq:dynamicS}). 
In Eq.~(\ref{eq:neutrinomb}), the integrand of the denominator is the free-space differential cross section per volume $V^{-1}d\sigma^\pm_0/dcos\theta$ defined similarly as in Eq. (\ref{eq:neutrinomb2}), but with the correction $S_A(q_0,q)$ replaced by $2\pi n$ where $n$ is the number density of neutrons (protons) for electron neutrino (electron anti-neutrino) reactions. The importance of nucleon-nucleon residual interaction in determining the neutrino-nucleon interactions was discussed in Ref.~\cite{PhysRevC.59.2888,Burrows:1998cg},  and the quantified correlation between $G'_0$ and many-body corrections was systematically studied in Ref.~\cite{PhysRevC.107.015804}. Comparing to Ref.~\cite{PhysRevC.107.015804}, in this work we improve the calculation of neutrino-nucleon many-body corrections by strictly solving the Bethe-Salpeter equation based on the method proposed in Ref.~\cite{Hernandez:1999zz,Margueron:2006wh}.

\begin{table*}[htb]
    \centering
\begin{tabular}{ccc}
\hline\hline
Skyrme Parameter& nuSkyI &nuSkyII\\
\hline
$t_0$ [$\mathrm{MeV~ fm^3}$] & -2585.9 & -2604.99\\
$t_1$ [$\mathrm{MeV~ fm^5}$]& 509.532  & 538.016 \\
$t_2$ [$\mathrm{MeV~ fm^5}$] & -68.789 & 25.3424\\
$t_3$ [$\mathrm{MeV~ fm^{3+3\alpha}}$]& 13071.1 & 13238.5\\
$x_0$ []& 0.456306 & 0.176233\\
$x_1$ []& -0.259363 & -0.212738\\
$x_2$ []& -0.579601 & -4.40879\\
$x_3$ []& 0.732344 &0.200248\\
$\alpha$ []& 0.129909 &0.13045\\
$b_4$ [$\mathrm{fm^4}$]& 0.261995 & 0.161697\\
$b'_4$ [$\mathrm{fm^4}$]& 0.270569 & 0.524227\\
\hline\hline
\end{tabular}
    \caption{ The Skyrme parameters of nuSkyI and nuSkyII.  }
    \label{tab:nuskyPara}
\end{table*}

\begin{table*}[htb]
    \centering
\begin{tabular}{llcccc}
\hline\hline
&&Experiment& nuSkyI &nuSkyII\\
\hline
&$BE/A$ [MeV]&~7.87 ~\cite{Wang:2012eof,Angeli:2013epw} & 7.73 & 7.89\\
&$R_{ch}$ [fm]&~5.50 ~\cite{Wang:2012eof,Angeli:2013epw}& 5.47  & 5.44 \\
{\bf$^{208}$Pb~~}&$F_{ch}$ []&~0.409~\cite{PREX:2021umo} & 0.412  & 0.417 \\
&$\Delta F$ []&~0.041 ~\cite{PREX:2021umo} & 0.0241  & 0.0288\\
&$E_\mathrm{GT}$ [MeV]&19.2 ~\cite{PhysRevC.85.064606,PhysRevC.52.604} & 20.04  & 19.46\\
&$m_0\%$ []&~$74\%$ ~\cite{PhysRevC.85.064606,PhysRevC.52.604} & $75.6\%$ & $74.8\%$\\
\hline
&$BE/A$ [MeV]&~8.355~\cite{Wang:2012eof,Angeli:2013epw} & 8.25 & 8.37\\
&$R_{ch}$ [fm]&~4.71~\cite{Wang:2012eof,Angeli:2013epw} & 4.69  & 4.67 \\
{\bf$^{132}$Sn~~}&$E_\mathrm{GT}$ [MeV]&13.97 ~\cite{PhysRevLett.121.132501} & 14.66  & 13.97\\
&$m_0\%$ []&~ $81\%$ ~\cite{PhysRevLett.121.132501}& $79.23\%$ & $78.72\%$\\
\hline
&$BE/A$ [MeV]&~8.71~\cite{Wang:2012eof,Angeli:2013epw} & 8.65  & 8.79\\
&$R_{ch}$ [fm]&~4.27 ~\cite{Wang:2012eof,Angeli:2013epw}& 4.23  & 4.22 \\
{\bf$^{90}$Zr~~}&$E_\mathrm{GT}$ [MeV]&~15.8 ~\cite{PhysRevC.55.2909} & 16.2 & 16.15 \\
&$m_0\%$ []&~$68\%$ ~\cite{PhysRevC.55.2909}& $69.8\%$ & $70.7\%$\\
\hline
&$BE/A$ [MeV]&~8.67~\cite{Wang:2012eof,Angeli:2013epw} & 8.78 & 8.86\\
&$R_{ch}$ [fm]&~3.48~\cite{Wang:2012eof,Angeli:2013epw} & 3.46  & 3.46\\
{\bf$^{48}$Ca~~}&$F_{ch}$ []&~0.158 ~\cite{CREX:2022kgg} & 0.155  & 0.155 \\
&$\Delta F$ []&~0.0277 ~\cite{CREX:2022kgg} & 0.0423  & 0.0422 \\
&$E_\mathrm{GT}$ [MeV]&~10.5 ~\cite{PhysRevLett.103.012503} & 11.0  & 10.8 \\
\hline
{\bf ~Landau-Migdal Parameter~~} & $G'_0$ []& $0.48\pm0.034$ (this work) & 0.35  & 0.41\\
\hline
&$R_{1.4}$ [km]&~ $12.71^{+1.14}_{-1.19}$ ~\cite{Riley:2019yda} ; $13.02^{+1.24}_{-1.06}$ ~\cite{Miller:2019cac} & 12.12 & 12.20\\
{\bf ~NS~~} & $M_{max}$ [$M_\odot$]& $>2.0$ & 2.08  & 2.05\\
\hline\hline
\end{tabular}
    \caption{The basic properties of nuclei and neutron stars predicted by the new Skyrme sets nuSky. In this Table, $\Delta F$ is the difference between the weak and charge form factors $F_{w}$ and $F_{ch}$. The fraction of GT strength $m\%$ is
    defined as the ratio of the area covered by the GTR spectrum in the intervals $[15,24]$, $[9,24]$ and $[12,22]$ MeV to the area covered by the spectrum below 31 MeV, for $^{208}\mathrm{Pb}$, $^{132}\mathrm{Sn}$ and $^{90}\mathrm{Zr}$ GTR spectrum respectively. The relative uncertainty of accurately measured values (the binding energy, the charge radii and the peak location of GTR of finite nuclei) are artificially chosen to be $2\%$ of the experimental values to ensure the feasibility of Monte Carlo simulations. 
    }
    \label{tab:nusky}
\end{table*}

\begin{figure*}
	\centering
	\includegraphics[width=0.90\textwidth]{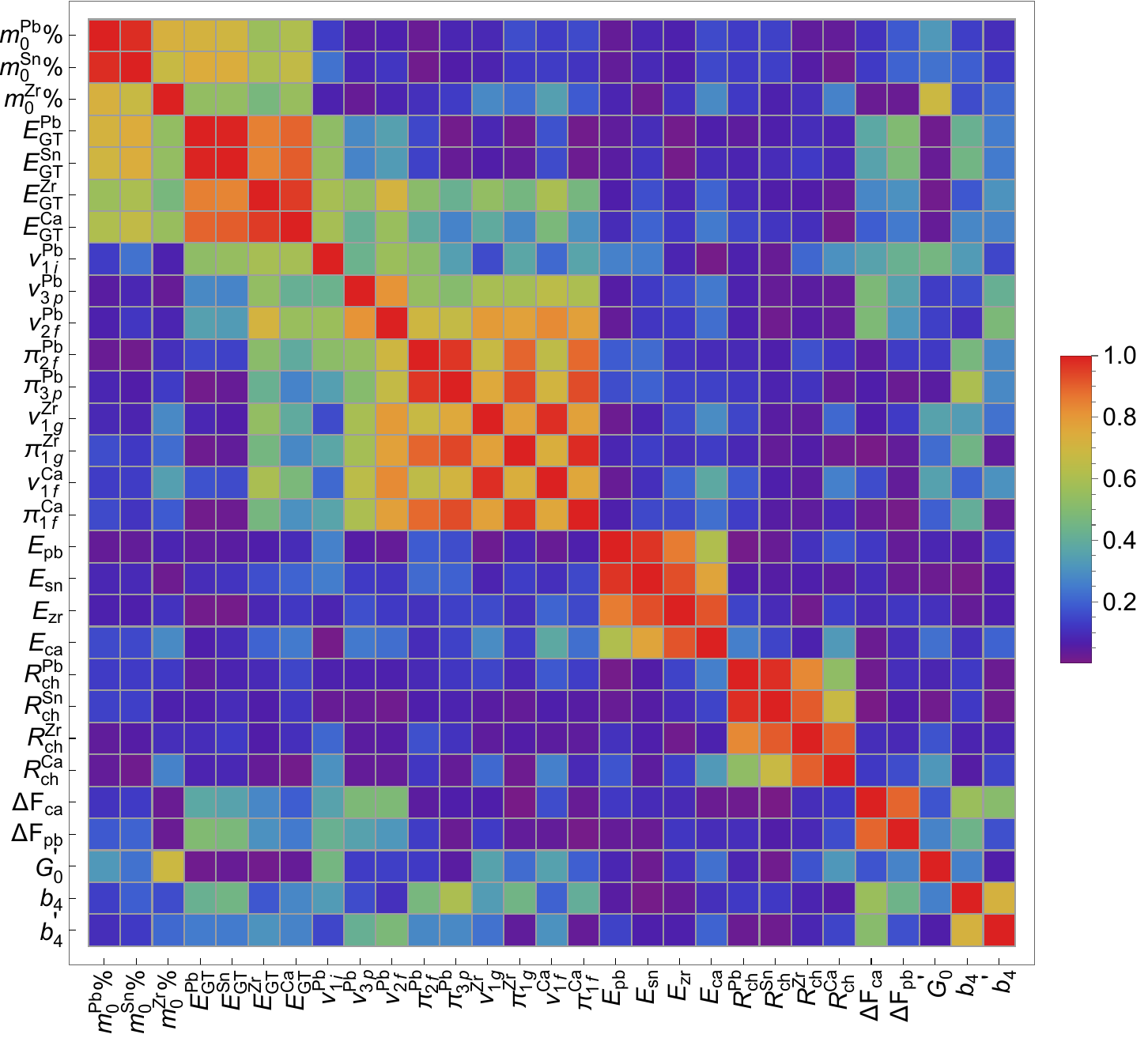}
	\caption{The Pearson correlation coefficients among the Gamow-Teller excited state features (e.g. $m_0\%$, $E_\mathrm{GT}$) and ground state features on various nuclei. The correlations of the sampled data are from the noGT calculations described in the text. } 
\label{fig:Peasonm0EGT1}	
\end{figure*}

\begin{figure*}
	\centering
    \includegraphics[width=0.90\textwidth]{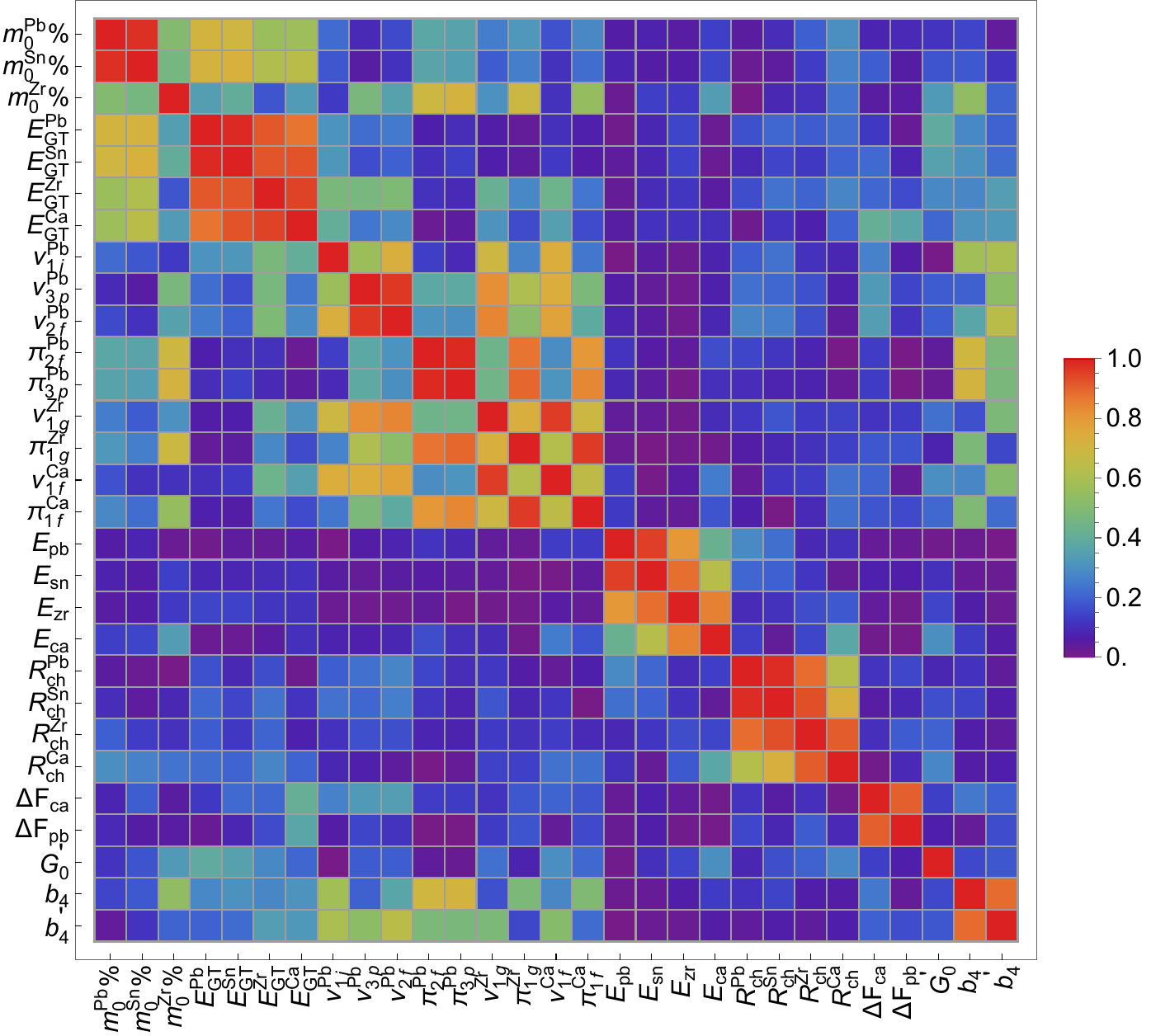}
	\caption{Similar as Fig. \ref{fig:Peasonm0EGT1}, but the correlations are from GTAll.} 
\label{fig:Peasonm0EGT2}	
\end{figure*}

\begin{figure*}
	\centering
	\includegraphics[width=0.45\textwidth]{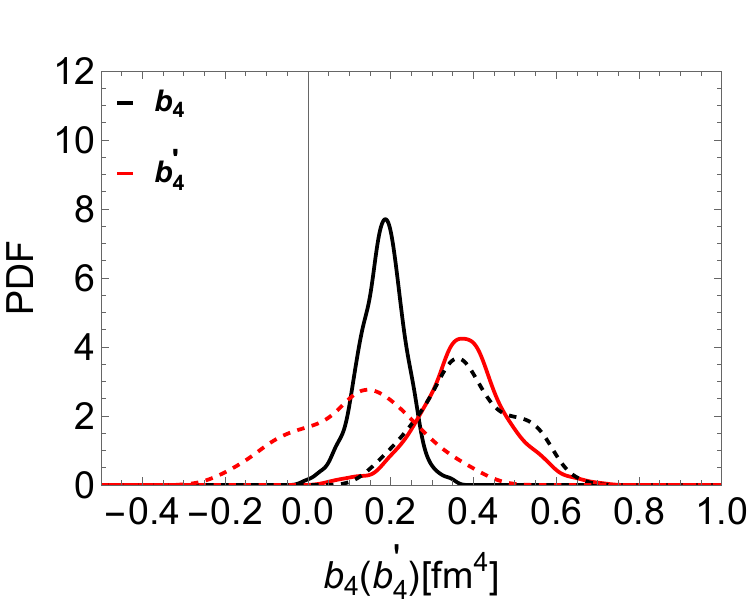}
	
	\caption{The PDF of $b_4$, $b'_4$ sampled from GTAll Bayesian inference (solid curves) and noGT inference (dashed curves). } 
\label{fig:b4b4p}	
\end{figure*}

\begin{figure}
	\centering
	\includegraphics[width=0.44\textwidth]{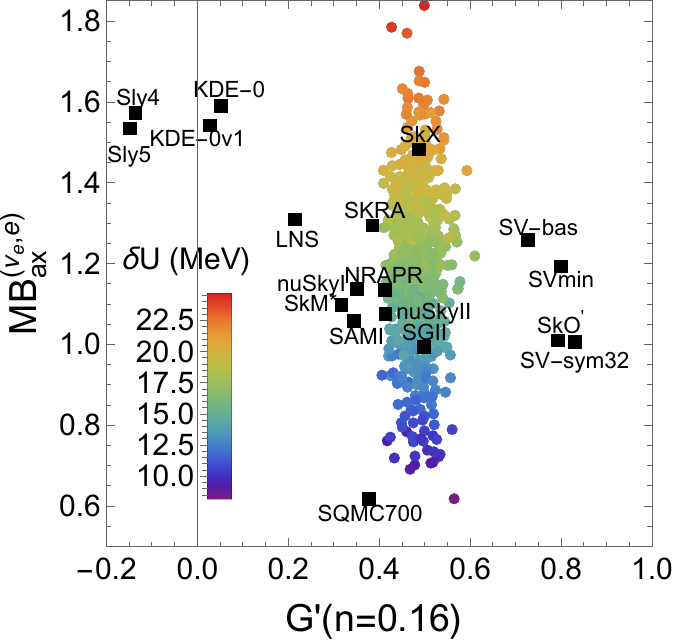}
	\includegraphics[width=0.455\textwidth]{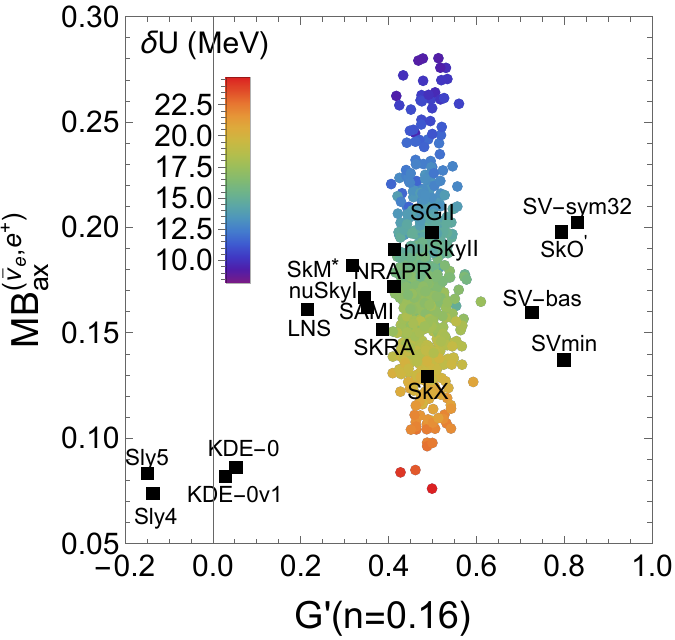}

	\caption{The dependence of neutrino-nucleon many-body (MB) corrections [defined in Eq.~(\ref{eq:neutrinomb})] on $G'_0$ from the GTAll inference described in the text. The black squares represent the MB corrections based on existing models. The upper panel shows the dependence of many-body corrections of $\nu_e(n,p)e^-$ on $G'_0$ and the lower panel shows the dependence of many-body correction of $\bar{\nu}_e(p,n)e^+$ on $G'_0$. The color represents the values of the difference of nucleon potentials $\delta U=U_n-U_p$, which is another factor determining the many-body corrections. The $\nu_e(n,p)e^-$, $\bar{\nu}_e(p,n)e^+$ reactions and the $\delta U$ examined here are calculated at $n=0.02~\mathrm{fm}^{-3}$, $T=10$ MeV, incoming neutrino energy $E_\nu=30$ MeV and $Y_e=0.05$. }
\label{fig:GpVsNeutrinoMB}	
\end{figure}

\section{Results} 

\subsection{Correlations in Bayesian Posterior}

In Fig. \ref{fig:Peasonm0EGT1} and \ref{fig:Peasonm0EGT2}, we present the Pearson correlation coefficients among the GTR features ($m_0\%$ and $E_\mathrm{GT}$), the spin-orbit splitting values and the ground-state properties of various nuclei. In the top (bottom) panel, the Pearson correlation coefficients are calculated based on posterior GTAll and noGT. The strength of Pearson correlations may or may not be sensitive to Bayesian constraints. 

By comparing the top and bottom panels, we find interesting correlations weakly-dependent on GTR constraints. An important observation is that the charge radii and the masses of investigated nuclei are very weakly correlated with the GTR features. This suggests that the Skyrme models have a good flexibility in describing the GTR features of finite nuclei, when the basic ground state properties of $^{48}$Ca, $^{90}$Zr, $^{132}$Sn, $^{208}$Pb of the Skyrme models are constrained. On the other hand, it suggests that, by only constraining the basic ground state properties of finite nuclei, the Skyrme models may not be able to give robust predictions of the GTR features. In addition, we observe that 1) $m_0\%$ and $E_{GT}$ of $^{208}$Pb (denoted as $m^{\mathrm{Pb}}_0\%$ and $E^{\mathrm{Pb}}_\mathrm{GT}$) are strongly correlated with $m_0\%$ and $E_{GT}$ of $^{132}$Sn; 2) the $E_\mathrm{GT}$ of $^{90}$Zr is strongly correlated with the $E_\mathrm{GT}$ of $^{48}$Ca; 3) the spin-orbit splitting of neutrons in the $3p$ channel is strongly correlated with the spin-orbit splitting of neutrons in the $2f$ channel of $^{208}$Pb (denoted as $\nu^\mathrm{Pb}_{3p}$ and $\nu^\mathrm{Pb}_{2f}$); 4) the spin-orbit splitting of protons in the $3p$ channel is strongly correlated with that of protons in the $2f$ channel in $^{208}$Pb (denoted as $\pi^\mathrm{Pb}_{3p}$ and $\pi^\mathrm{Pb}_{2f}$); 5) the spin-orbit splitting values of $^{48}$Ca and $^{90}$Zr are strongly correlated; 6) the ground-state properties (e.g. binding energy, charge radii) of $^{208}$Pb and $^{132}$Sn are strongly correlated; 7) the ground-state properties of $^{90}$Zr and $^{48}$Ca are strongly correlated; 8) the ground-state properties of $^{208}$Pb and $^{132}$Sn are correlated with those of $^{90}$Zr and $^{48}$Ca, but with less correlation strength. These obvious correlations appear as diagonal blocks in the figure. 

We also observe correlations sensitive to GTR constraints. Firstly, the correlation between $m_0\%$ and $G'_0$ is more obvious in the noGT models, which can be easily understood. In GTAll, $m_0\%$ is severely restricted, resulting in a weakening of the correlation. Indeed, if $m_0\%$ is totally fixed, then no correlation involving $m_0\%$ can be found. Secondly, the spin-orbit splitting values are correlated with the Skyrme parameters $b_4$ and $b'_4$. This correlation seems more obvious in GTAll. 

It is also worth mentioning the interesting correlation between $b_4$ and $b'_4$. Historically, $b'_4$ (which tunes the isovector spin-orbit force) was introduced as a parameter independent from  $b_4$ to increase the flexibility of the predictions for the charge radii along isotopic chains \cite{Reinhard:1995zz}. However, as shown in Fig. \ref{fig:Peasonm0EGT2}, the correlation between $b_4$ and $b'_4$ is significantly strengthened when GTR constraints are applied. This suggests that simultaneously fitting the charge radii over a wide region of the nuclear chart and fitting the GTR features of finite nuclei based on existing Skyrme models may be challenging. This potential tension should be investigated in future works.

In addition, we report the posterior distribution of the two spin-orbit Skyrme parameters $b_4$ and $b'_4$. These two parameters have an influence on the charge radii \cite{Reinhard:1995zz} and neutron skin thickness \cite{Zhao:2024gjz} of finite nuclei. They are also related to spin-orbit splitting and GTR properties. In Fig. \ref{fig:b4b4p}, we present the posterior of $b_4$ and $b'_4$ in two Bayesian inferences: one is based on noGT (dashed curves) and the other is based on GTAll (solid curves). The inclusion of GTR constraints significantly changes the distribution of these two Bayesian parameters. The posterior of the latter suggests that $b'_4>b_4>0$. Interestingly, the GTR data suggests relatively large $b'_4$ values, which coincides with the recent findings based on joint PREX and CREX analysis \cite{Zhao:2024gjz,Yue:2024srj}. Although very large $b'_4$  might be in tension with the shell structures \cite{Kunjipurayil:2025xss}, the large ensemble of Skyrme parameters sampled from GTAll can be used to systematically estimate the strength of this tension and search for models mitigating this tension. %

\subsection{nuSky models}

After investigating the correlations and the tensions in GTAll, we propose two new Skyrme sets (nuSkyI and nuSkyII in Tab.~\ref{tab:nuskyPara}) found in GTAll, which reasonably describe the ground state and Gamow-Teller resonance properties of various finite nuclei and also basic neutron star properties (as shown in Tab. \ref{tab:nusky}). 
The symmetry energy slope of nuSkyI and nuSkyII is $L=53.4~ \mathrm{MeV}$ and $L=64.2~\mathrm{MeV}$,  respectively. This new Skyrme parameterization series (nuSky) may provide a self-consistent description of the macroscopic massive star properties (such as the neutron star mass and radii) and the microscopic weak interaction properties in the stars (such as neutrino-nucleon and neutrino-nucleus interactions in hot and dense matter). 

Both nuSkyI and nuSkyII have reasonable infinite dense matter bulk properties from low to high densities, so that they can be used to describe neutron star properties. They are required to fulfill the following high-density properties: 1) the effective mass of protons and neutrons up to $1.0 ~\mathrm{fm^{-3}}$ are non-negative; 
2) the speed of sound up to $2.0~\mathrm{fm^{-3}}$ is positive; 3) the beta equilibrium can be achieved up to $2.0~\mathrm{fm^{-3}}$; 4) the di-neutron state does not appear below saturation density by requiring that the energy per nucleon is always positive in pure neutron matter. The nuSkyI has the highest total likelihood among a subset of the models with small intrinsic scattering parameters which corresponds to a systematic uncertainty $<2\%$. This model can be approximately considered as an optimal model based on Bayesian inference without considering the systematic uncertainty. The nuSkyII has the highest total likelihood among all the parameterizations that satisfy the aforementioned high-density neutron star matter properties.

\subsection{Neutrino Opacity}
In Fig.~\ref{fig:GpVsNeutrinoMB}, we present the statistical distribution of $\mathrm{MB}_{ax}$ based on the GTAll posterior, and compare it with those based on existing Skyrme parameterizations (black squares). The distribution based on existing Skyrme models and the study in Ref.~\cite{PhysRevC.107.015804} indicate that $\mathrm{MB}_{ax}^{(\nu_e,e^-)}$ ($\mathrm{MB}_{ax}^{(\bar{\nu}_e,e^+)}$) decreases (increases) with the increase of $G'_0$. For the first time, by applying GTR constraints we significantly reduce the uncertainty of $\mathrm{MB}_{ax}$ due to variation of $G'_0$. Indeed, at densities around the neutrino sphere, large positive $G'_0$ may neutralize or even reverse the mean-field effect that enhances (suppresses) the electron (anti)neutrino opacity \cite{PhysRevC.107.015804,Shin:2023sei,Burrows:1998cg}. The remaining uncertainty of $\mathrm{MB}_{ax}$ mainly comes from the variation of $\delta U=U_n-U_p$ at sub-saturation densities and results in a wide spreading of $\mathrm{MB}_{ax}$ along the vertical direction. The neutron skin measurements ~\cite{PREX:2021umo,CREX:2022kgg} applied in our Bayesian inference cannot tightly constrain the symmetry energy and $\delta U$ at sub-saturation densities. %

The virial equation of state \cite{Horowitz:2012us} and the chiral effective field theory calculations~\cite{Rrapaj:2014yba} may provide robust predictions of this quantity. In addition, the cluster formation around the neutrino sphere results in non-uniform nucleon density distributions and may have a significant influence on $\delta U$. All these interesting questions related to the nucleon potentials in hot nuclear matter at sub-saturation densities are left for future studies.

\section{Summary}  

In this work, for the first time, we provide an extensive statistical analysis of the correlations emerging from the Skyme models based on GTall and noGT ensembles. The quantified correlations of those finite-nuclei properties measure the ability of standard Skyrme EDFs to simultaneously describe both ground-state and GTR features. As expected, the GTR features and the ground-state properties of finite nuclei are weakly correlated within this theoretical framework. On the other hand, the GTR properties of the representative closed shell nuclei investigated in this work, are obviously correlated.  

Based on a large ensemble of Skyrme models sampled from GTAll, we propose two new Skyrme energy density functionals nuSkyI and nuSkyII, which successfully describe the GTR features, the spin-orbit splitting, the basic ground state properties of selected nuclei and the neutron star properties. The nuSky models may provide a self-consistent description of the macroscopic properties of compact stars, as well as the microscopic weak reactions inside the compact stars.  

Finally, we compute the many‑body corrections on axial‑current neutrino–nucleon absorption reactions in hot and dense matter using the GTAll ensemble of Skyrme interactions. These interactions are constrained simultaneously by ground‑state observables and Gamow–Teller resonance data, and their statistical uncertainties can be quantitatively propagated to the charged‑current neutrino opacities in core‑collapse supernovae and binary‑neutron‑star mergers. This analysis substantially reduces one of the major uncertainties in neutrino‑opacity theory—namely, the treatment of nucleon–nucleon residual interactions in the spin–isospin channel. In addition to the spin‑dependent residual interactions, we find that the symmetry energy below saturation density also plays an important role in shaping neutrino–nucleon interactions; this quantity could be further constrained by virial equation of state and chiral EFT calculations. A systematic study of how residual interactions and the symmetry energy jointly influence neutrino opacities across a wide range of densities, temperatures, and proton fractions will require future works. 

\subsection*{Acknowledgement}

ZL and AWS were supported by NSF PHY 21-16686. AWS was also supported by the Department of Energy Office of Nuclear Physics. This work used Bridges2 at University of Pittsburgh through allocation PHY230061 from the Advanced Cyberinfrastructure Coordination Ecosystem: Services $\&$ Support (ACCESS) program, which is supported by U.S. National Science Foundation grants $\#2138259$, $\#2138286$, $\#2138307$, $\#2137603$, and $\#2138296$.

\bibliography{References}
\bibliographystyle{apsrev} 

\end{document}